\documentclass[aps,pre,twocolumn]{revtex4-1}
\usepackage{amssymb}
\usepackage{amsmath,bm}
\usepackage{graphicx,color}
\usepackage{bbm}
\usepackage{multirow}

\DeclareMathAlphabet{\mathitbf}{OML}{cmm}{b}{it}

\newcommand{\B}[1]{{\bm{#1}}}
\newcommand{\C}[1]{{\mathcal{#1}}}    
\let \= \equiv \let\*\cdot \let\~\widetilde \let\^\widehat \let\-\overline

\begin{document}
\title{Diffusion in Agitated Frictional Granular Matter Near the Jamming Transition}
\author{H.G.E. Hentschel$^{1,2}$, Itamar Procaccia$^1$ and Saikat Roy$^1$}
\affiliation{$^1$Dept. of Chemical Physics, The Weizmann Institute of
Science, Rehovot 76100, Israel
\\$^2$Dept of Physics,
Emory University, Atlanta, Georgia, }

\begin{abstract}
We study agitated frictional disks in two dimensions with the aim of developing a scaling theory
for their diffusion over time. As a function of the area fraction $\phi$ and mean-square velocity
fluctuations $\langle v^2\rangle$ the mean-square displacement of the disks $\langle d^2\rangle$ spans
4-5 orders of magnitude. The motion evolves from a subdiffusive form to a complex diffusive behabvior at long times. The statistics of $\langle d^n\rangle$ at all times are multiscaling, since the probability distribution function (pdf) of displacements has very broad wings. Even where a diffusion constant can be identified it is
a complex function of $\phi$ and $\langle v^2\rangle$. By identifying the relevant length and time scales and their interdependence one can rescale the data for the mean square displacement and the pdf of displacements into collapsed scaling functions for all $\phi$ and $\langle v^2\rangle$. These scaling functions provide a predictive tool, allowing to infer from one set of measurements (at a given $\phi$ and $\langle v^2\rangle$) what are the
expected results at any value of $\phi$ and $\langle v^2\rangle$.
\end{abstract}
\maketitle
\section{Introduction}

The term ``Frictional Granular matter" pertains to macroscopic solid granules which interact via {\em normal} forces due to compressional contacts and {\em tangential} forces that are due to mutual friction. Upon collisions the restitution coefficient is often smaller than unity. Thus in the absence of forcing, shaking,  or any other manner of excitation such matter attains mechanical equilibrium where the net force (and torque) on every granule is zero. Continuous dynamics in such matter can be induced by agitation (in a sense made precise below), such that ``life" is given to the granules that may begin to diffuse around the sample. In this paper we are motivated by experiments of the type reported in Refs.~\cite{05KMPKP,05CKLLC,06AD,07KAGD,08LDBB,08ZBGPB,10KK} except that we are less concerned with the jamming criticality and more with the unusual diffusion dynamics. To this
aim we consider below ``temperatures" that are typically higher than those considered for example in Ref.~\cite{08LDBB}, such that the diffusive process becomes less sensitive to the zero-temperature jamming criticality. The ``temperature" is defined below by maintaining a chosen level of velocity variance $\langle v^2\rangle$; to this aim we employ stochastic random kicks, and see below for details.

The statistics of the diffusion process can be characterized by the moments of the time-displacements of the disks. Denoting by $\B r_i$ the coordinate of the center of mass of the $i$th disk in a system with $N$ disks, the displacement over time
is written as
\begin{equation}
\B d_i(t)\equiv \B r_i(t+s) - \B r_i(s) \ ,
\end{equation}
where time translational invariance has been assumed. In two dimension we can write (cf. Fig.~\ref{figpdf} below)
$d_i^2 = d^2_{i,x}+d^2_{i,y}$. In this paper we employ a number of moments of this displacement:
\begin{eqnarray}
\langle d^2_{i,x}\rangle &=&\langle d^2_{i,y}\rangle \equiv  \langle [r_{i,x}(t+s) - r_{i,x}(s)]^2\rangle\ ,\\
\langle d^2\rangle(t) &\equiv& \frac{1}{N}\sum_{i=1}^N\langle [\B r_i(t+s) - \B r_i(s)]^2\rangle \ ,\\
\langle d^n\rangle(t) &\equiv& \frac{1}{N}\sum_{i=1}^N\langle [\B r_i(t+s) - \B r_i(s)]^n\rangle \ , \quad n=3,4,\cdots \ . \nonumber
\label{defdn}
\end{eqnarray}
Here $\langle \cdots \rangle$ stands for a time average over $s$. Space isotropy guarantees that all the odd
moments vanish, and therefore below we consider only even moments. When the area fraction $\phi$ is close
to the jamming point of the un-agitated system these dynamics are highly heterogeneous both in space and in time.
In space one recognizes regions of the system that are quite motionless due to local jamming and only a few particles contribute to the averages in Eq.~(\ref{defdn}). In addition, caging will insure temporal heterogeneity
due to pinning of the disks in a given cage for long periods of time with rare ballistic spurts of motion.
In consequence these moments
display complex (multiscaling) statistics which we expose below.

The stress in this paper is on the second moment $\langle d^2\rangle(t)$ whose time dependence exhibits
a crossover from subdiffusive to diffusive behavior. We stress at this point, that the fact that the second moment grows like time for large times does not imply simple scaling for higher moments. ``Diffusive behavior" is understood throughout this paper to have only this meaning.  Elucidating the full dependence of $\langle d^2\rangle(t)$ on time, on $\phi$ and on $\langle v^2\rangle$ is rather demanding, and it calls for a careful identification of
the crucial energy scales, length scales and time scales dominating these dynamics. These scales turn out to be
the energy barrier $\Delta(\phi)$ for disk hopping, the diverging length scale $\xi(\phi)$ associated with the
distance between diffusing regions (and see below) and the crossover time scale $t_\xi$ between subdiffusion and diffusion. Once the data
for the diffusion process is properly rescaled we find data collapse that can be used to predict how
diffusion occurs at any value of $\phi$ and $\langle v^2\rangle$ and time $t$ from the measurement
of one set of these parameters.

The structure of the paper is as follows: in Sect.~\ref{model} we set up the numerical simulations and explain
the dynamics of the agitated amorphous assembly of disks. In Sect.~\ref{simresults} we present the results
of numerical simulations and motivate the scaling theory that comes next. Section \ref{scaling} is the central
section in this paper, and it details the identification of the relevant scale and elaborates on their
use in a scaling theory. In Sect.~\ref{compare} we reap the benefit of the insights obtained in Sect. \ref{scaling}, and provide the scaling functions that can be compared directly with the results of
numerical simulations. We find excellent data collapse and an a-posteriori justification to our scaling
assumptions. Section \ref{discussion} offers a discussion of some of the salient assumptions about different
length scales, a summary of the paper and some conclusions.

\section{Setting up the numerical experiments}
\label{model}

\subsection{Model forces}
\label{forces}
The model studied below employs a binary assembly of $N$ frictional disks of mass $m$ in a two-dimensional box, half of which with radius $\sigma_1=0.5$ and the other half with $\sigma_2=0.7$. The area fraction $\phi$ is defined as the ratio of the nominal area occupied by the disks divided by the area of the box. Without agitation there exists an area fraction $\phi_J$ such that for $\phi<\phi_J$ the disks are not interacting and the pressure is zero. Under external stress our disks interact with binary interactions; the normal
force is determined by the overlap $\delta_{ij} \equiv \sigma_i+\sigma_j-r_{ij}$ where
$\B r_{ij}\equiv \B r_i-\B r_j$. The normal force is Hertzian, but we allow collisions to be inelastic. We therefore write
\begin{equation}
\B F_{ij}^{(n)} = K_n \delta_{ij}^{3/2}\hat r_{ij} -\gamma_n |(\dot{\B  r_i}-\dot{\B  r_j})\cdot \hat r_{ij}|\hat r_{ij}\ , \quad \hat r_{ij} \equiv \B r_{ij}/r_{ij} \ ,
\label{Fn}
\end{equation}
with $K_n=2\times 10^5$. The mass of disks are $m=1$ and the units below will be determined by
$m$, $2\sigma_1$ and time in units of $1/\sqrt{K_n}$.

The tangential force takes into account the tangential displacement $\B t_{ij}$. Upon first contact between two disks $t_{ij}=0$. Providing every disk with the angular coordinate $\theta_i$ we can
accumulate tangential stress according to
\begin{equation}
d\B t_{ij} =d\B r_{ij} -(d\B r_{ij}\cdot \B r_{ij}) \hat r_{ij} +\hat r_{ij} \times (\sigma_i d\theta_i +\sigma_jd\theta_j) \ .
\end{equation}
The Mindlin model of the tangential force is \cite{49Min}
\begin{equation}
\B F_{ij}^{(t)} = -K_t\delta_{ij}^{1/2}t_{ij} \hat t_{ij} \ ,
\label{Min}
\end{equation}
together with the Coulomb condition
\begin{equation}
\B F_{ij}^{(t)} \le \mu \B F_{ij}^{(n)} \ ,
\label{Coul}
\end{equation}
where $\mu$ is the friction coefficient. Below we use $K_t=K_n$ and $\mu=0.1$.

The initial conditions for the simulations are obtained by starting with a random placement of small and large
disks in a rectangular box of dimension $57\sigma \times 102\sigma$, large enough for the pressure to be zero.
The system is then compressed quasistatically to a final box with a chosen value of the area fraction $\phi$.

\subsection{Dynamics}
\label{dynamics}

The dynamics is provided by the second order equations for the coordinates $\B q_i\equiv r_i^x, r_i^y, \theta_i$
for $i=1,2\dots N$. Without additional global damping or agitation the set of equations read
\begin{equation}
m_i\frac{d^2 \B r_i}{dt^2} =\B F_i \ , \quad I_i\frac{d^2 \theta_i}{dt^2} = \sigma_i\hat r_{ij} \times F^{t}_{ij} \ .
\end{equation}
where $I_i$ are the moments of inertia. Below we add random kicks to the translational degrees of freedom compensated by an additional background damping term. In addition, to bring the model closer to experimental reality, we allow a restitution coefficient smaller than unity. Thus energy is not conserved and as a consequence ``life" is provided only by the random kicks. We are dealing with an open system and  cannot expect fluctuation-dissipation relations to hold as in thermal equilibrium \cite{07PBV}. Note that with these dissipative contributions the equation for the  angular degree of freedom is not changed. In other words we write
\begin{equation}
m_i\frac{d^2 \B r_i}{dt^2} =\B F_i -m_i\gamma \frac{d\B r_i}{dt} +\B f_i(t) \ ,
\end{equation}
with $\B f_i(t)$ being a $\delta$-correlated random force with zero mean:
\begin{equation}
\langle \B f_i(t) \cdot \B f_j(t+\tau)\rangle = 2 \Gamma \delta (\tau) \delta_{ij}
\end{equation}
 For the numerical implementation we assume the time scale for the random kicks is very short compared to the deterministic terms but otherwise continuous. We thus replace the above stochastic equation with a Stratonovitch stochastic differential equation with an inner time scale denoted by $dt$. Thus
\begin{equation}
\label{strat}
\B f_{i}(t) =  \sqrt{ \frac{\Gamma}{dt}} \B R_{i}(t) \ ,
\end{equation}
where $ \B R_{i}(t)$ is a bounded random noise  with zero mean and unit variance
\begin{eqnarray}
\label{stratf}
\langle  \B R_{i}(t) \rangle & = & 0 \nonumber \\
\langle \B R_i(t)\cdot \B R_i(t)\rangle & = & 1
\end{eqnarray}

At this point we choose in our numerics a value of $\Gamma$ to achieve (in the steady state) a desired
pre-determined value of the velocity variance $\langle v^2\rangle$. For notational convenience we
employ an effective temperature $T$ in units in which the Boltzmann constant equals
unity. By definition this temperature is
\begin{equation}
\label{temp}
T= m\langle v^2\rangle/2 \ .
\end{equation}
If we want a given value of $T$ when we change $\phi$ we need to change $\Gamma$ to achieve the same level of $\langle v^2\rangle$. At higher packing fractions there is increased
dissipation and therefore the average steady-state kinetic energy decreases. In other words
all the temperatures below are reported in terms of $m \langle v^2\rangle/2$.

\section{Grain Dynamics close to the Jamming Transition}
\label{simresults}

\begin{figure}
\includegraphics[scale=0.2]{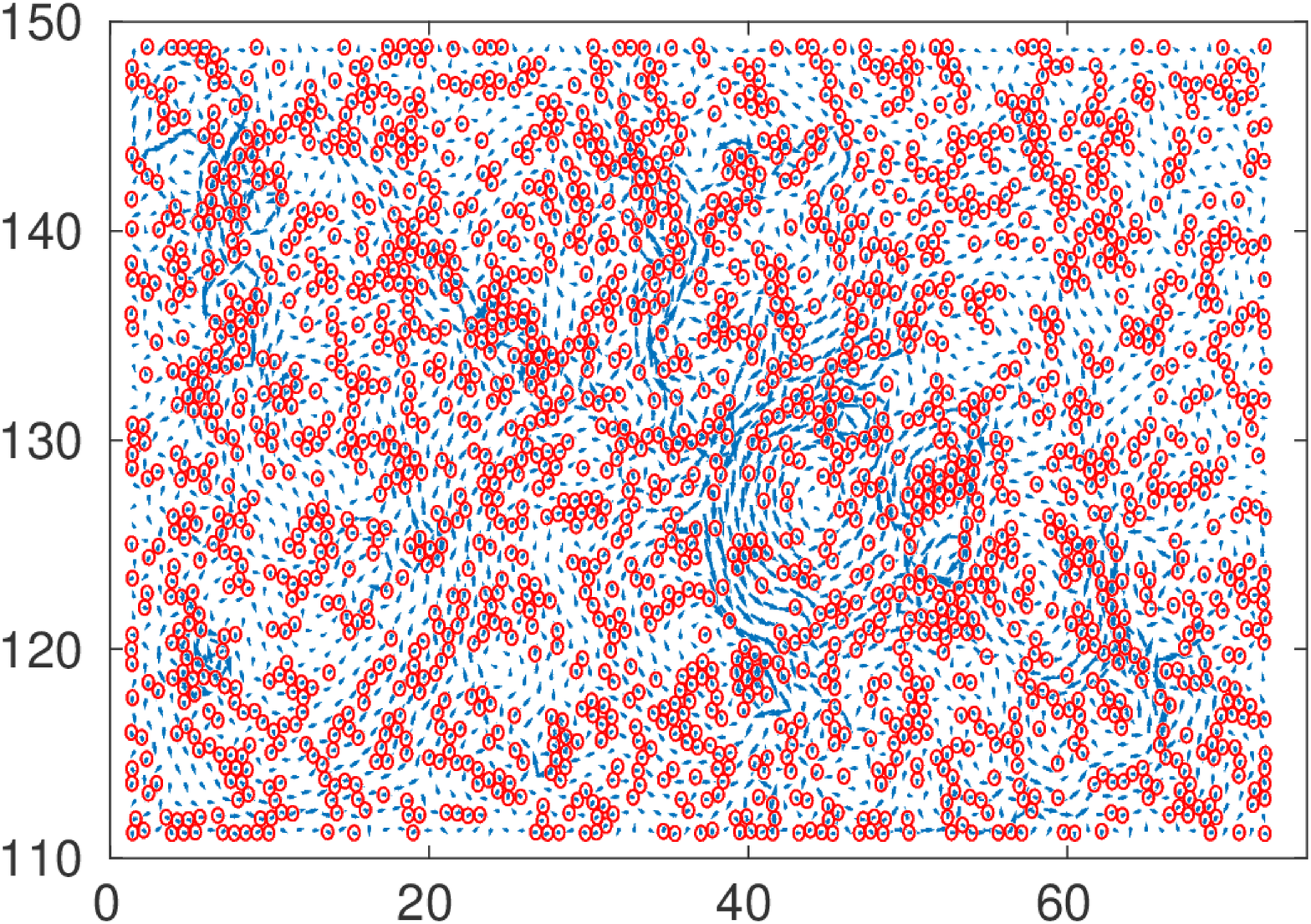}
\vskip 0.1 cm
\includegraphics[scale=0.2]{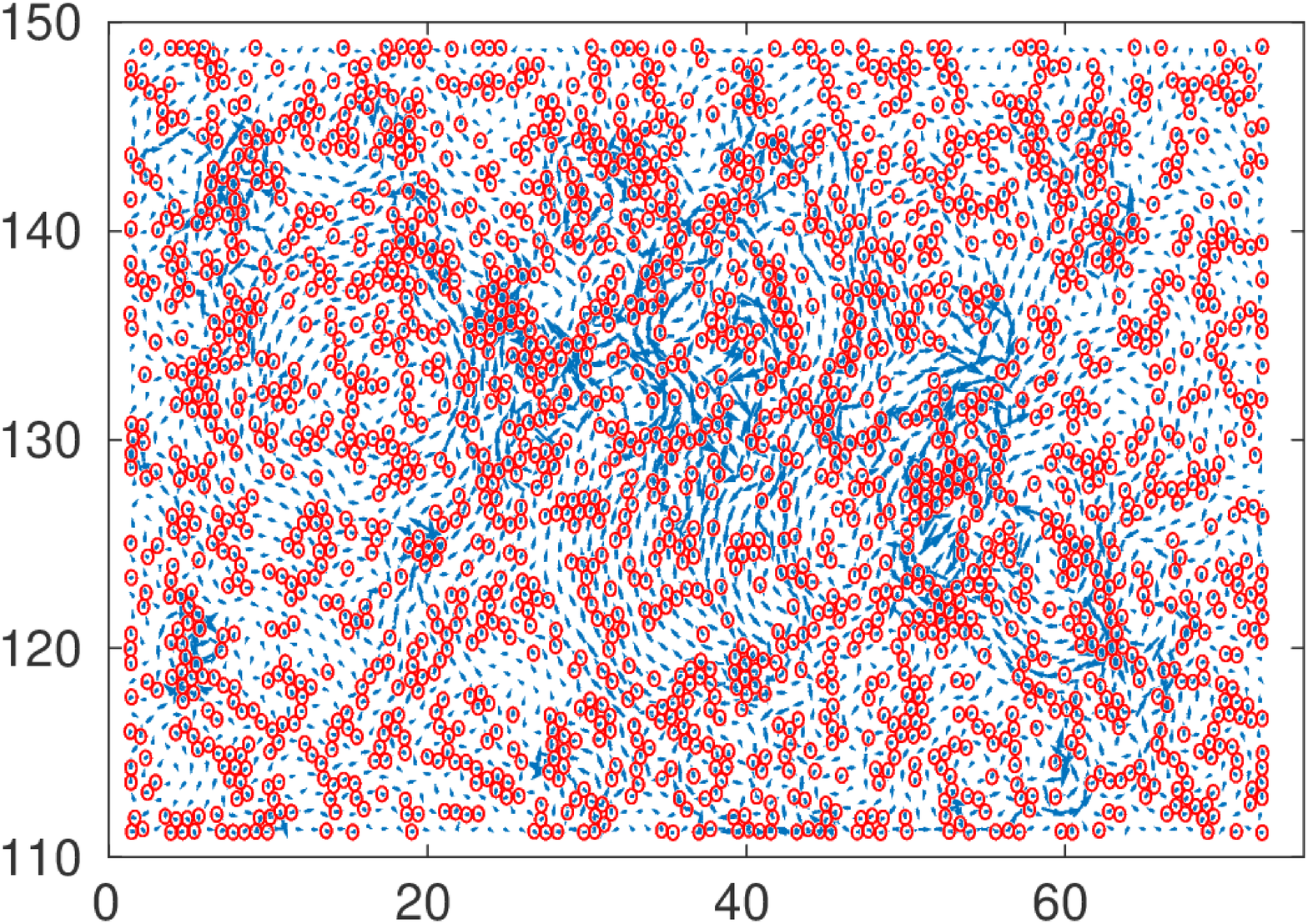}
\caption{ Upper panel: Total motion of all grains plotted after $5.4\times10^5$ time steps. For presentation purposes here and in Fig.~\ref{particlemotionb} we show as red disks only the small particles. Using blue arrows we indicate the displacements of both small and large particles. Lower panel: Total motion of all grains plotted after $2\times10^6$ time steps. Both displacement
fields are for $\phi=0.8319$ which is close to jamming area fraction which is estimated below to be $\phi_J\approx 0.838$. }
\label{particlemotion}
\end{figure}

Before making any scaling ansatz let us examine what simulations tell us about the statistics of displacements.
Simulations suggest several points:
\begin{itemize}
\item {\bf Diffusion is spatially heterogenous.} This can be seen in figures~\ref{particlemotion} and~\ref{particlemotionb}.  Here are plotted the trajectories of all particles for the two times $t=5.4\times10^5$ and $t=2\times10^6$ and two different packing fractions $\phi=0.822$ which is certainly in the granular solid regime but well below the jamming fraction $\phi_J$, and for $\phi=0.8319$ which is closer to the jamming fraction $\phi_J\approx 0.838$.  The visualization take home message is that (at least at short times) the disks split into two sub populations, those that diffuse appreciably and those that
are basically static. Note that as the jamming fraction $\phi_J$ is approached the heterogenous regions that are displaced become ever sparser and at the same time involve fewer particles that are displaced by significant amounts. It is important to realize, however, that even the ``significant displacements" are only a few $\sigma$. It is clear that diffusion must begin in regions where the local distribution of grains are loosely packed compared to the average particle density, defining  nucleating points for diffusion. These nucleating points for diffusion are separated by some length scale $l_{\rm nuc}(\phi )$ which depends on the volume fraction $\phi$. We expect this typical length to increase as $\phi$ does. Note that the region explored by diffusing disk which surrounds these nucleating sites grows also with time $t$;  we denote the size of this growing regions as $R(t)$, allowing ever more grains to move in a cooperative fashion.
\begin{figure}
\includegraphics[scale=0.20]{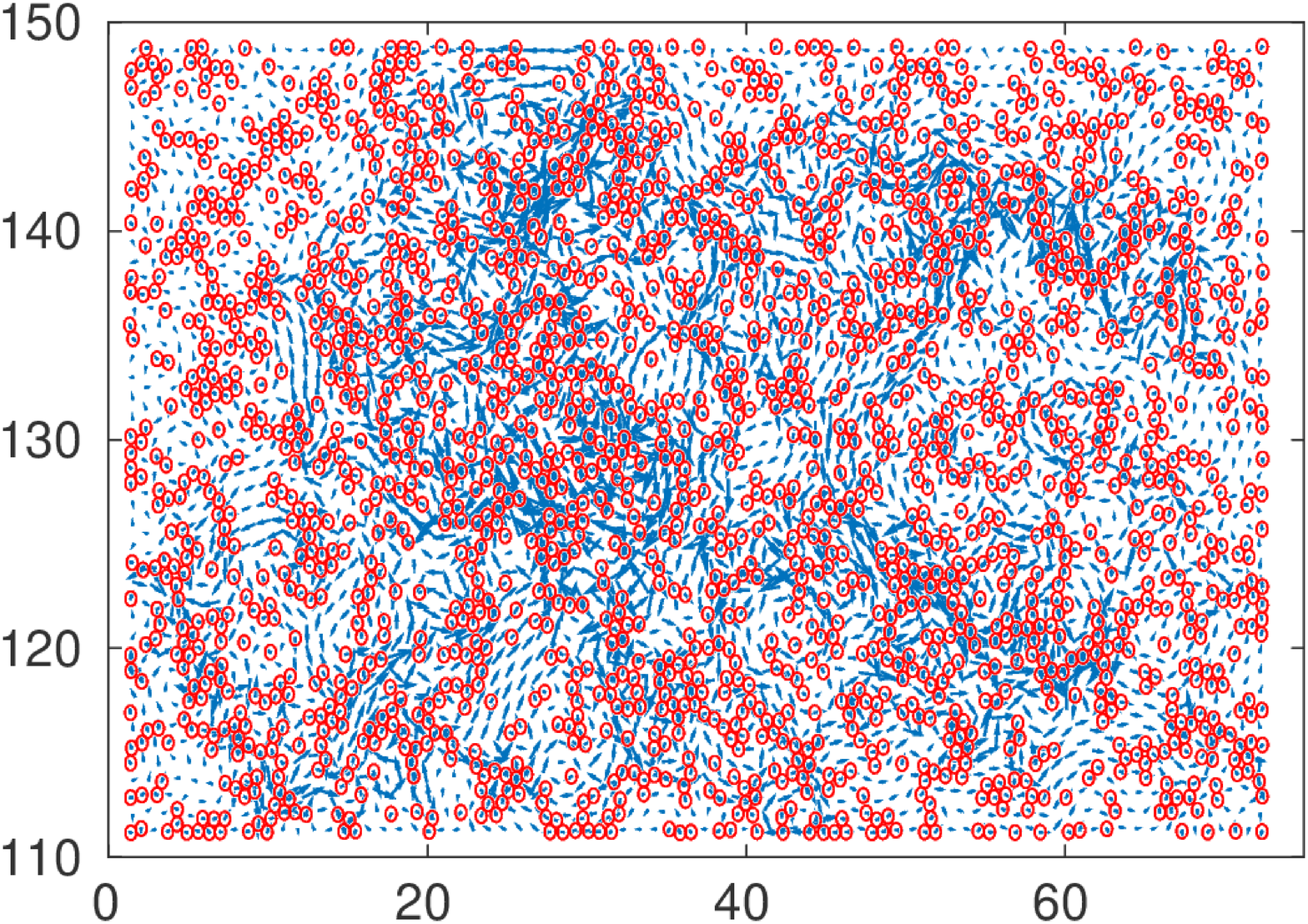}
\vskip 0.1 cm
\includegraphics[scale=0.20]{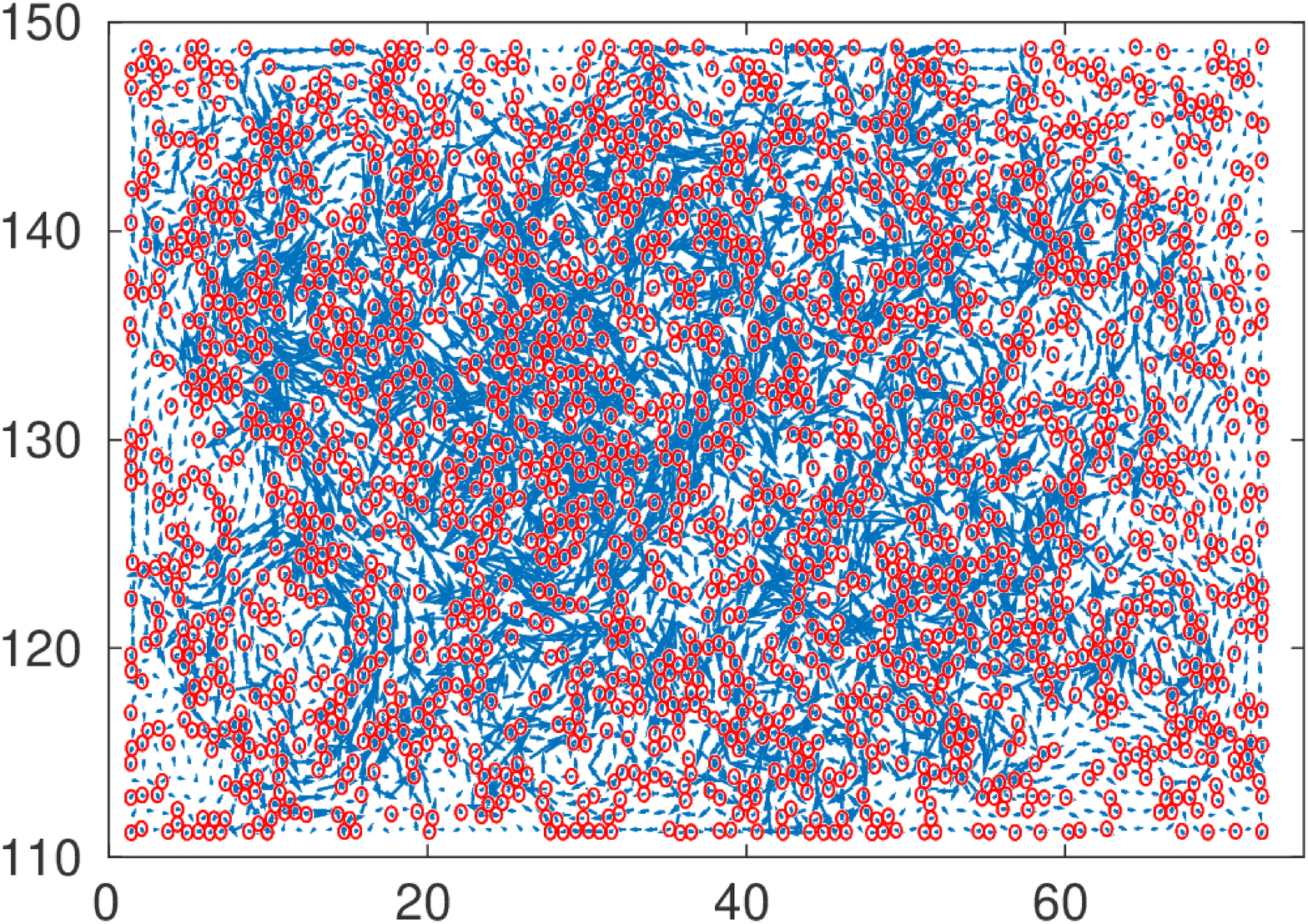}
\caption{ Upper panel: Total motion of all grains plotted after $5.4\times10^5$ time steps. Lower panel: Total motion of all grains plotted after $2\times10^6$ time steps. Both displacement
fields are for $\phi=0.822$ a packing fraction which is farther from $\phi_J$ . }
\label{particlemotionb}
\end{figure}

\item {\bf The motion is temporally heterogenous} --many grains hardly move while others appear to perform intermittent flights between cages, cf. Fig.~\ref{sdt}. Between flights  the particles can be stuck for long periods of time. Note again the spatial and temporal scales involved. The intermittent flights are occurring on very small scales involving often less than a single grain scale. Even the largest displacements involve just a few grain diameters.
\begin{figure}
\includegraphics[scale=0.18]{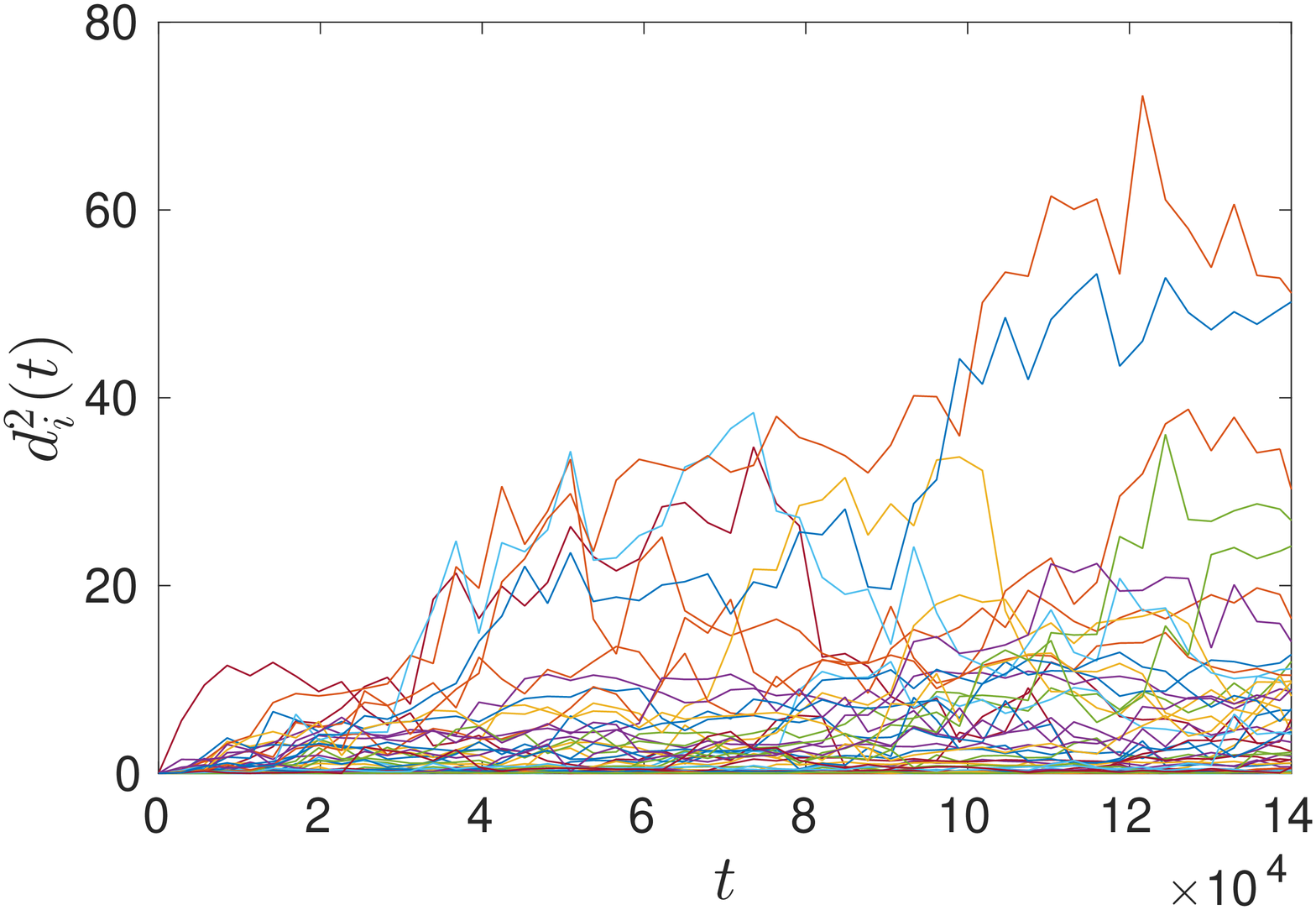}
\vskip 0.1 cm
\includegraphics[scale=0.18]{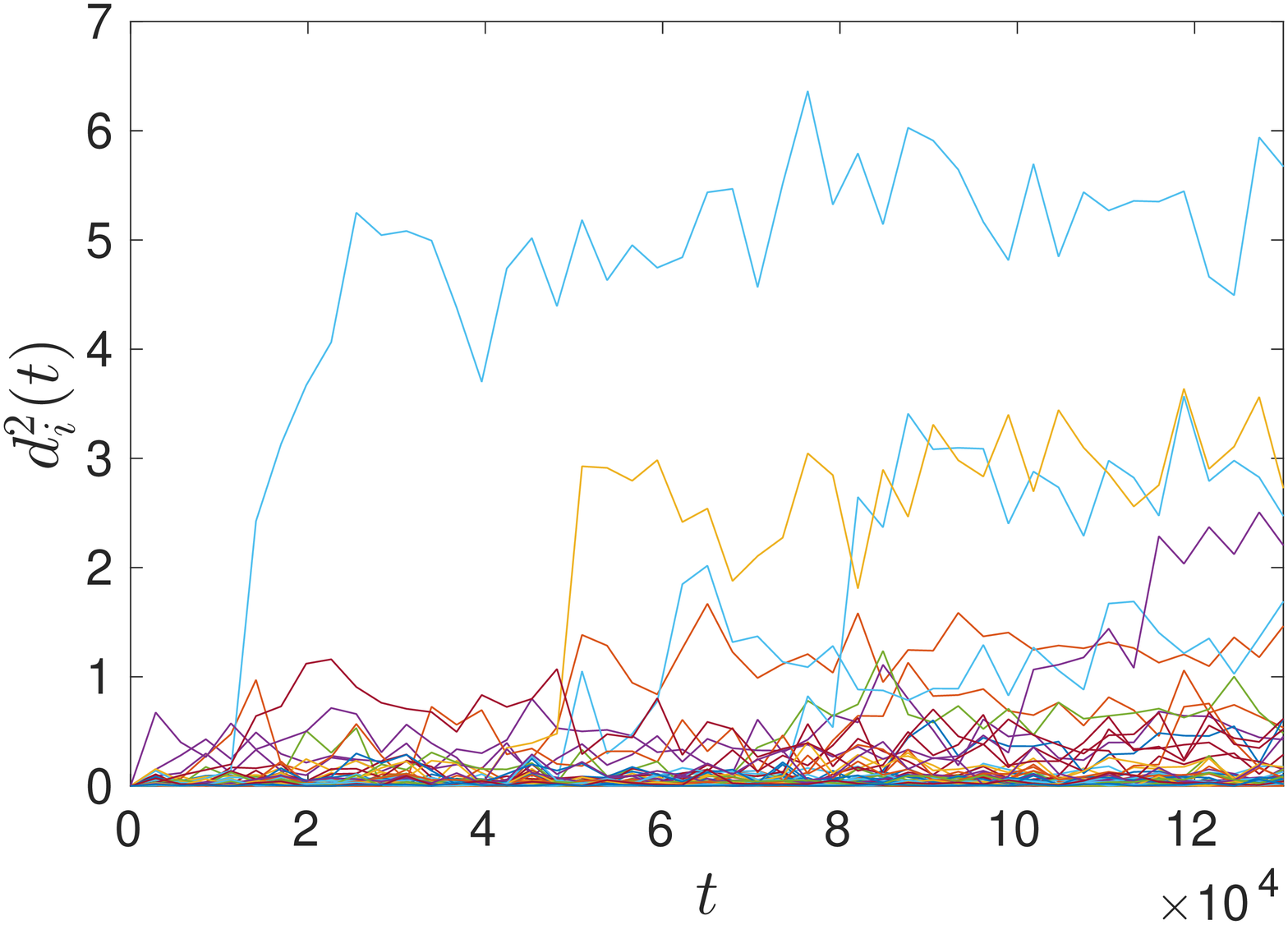}
\caption{ Upper panel: Squared displacement versus time for $50$ randomly chosen particles at packing fraction $\phi=0.822$. Lower panel: Same quantities at packing fraction $\phi = 0.8319$. Note the small number of
disks that diffuse significantly and how this number reduces when $\phi$ increases.}
\label{sdt}
\end{figure}

\item {\bf Simulations suggest a crossover time scale $t_\xi(\phi,T )$ between subdiffusive and diffusive dynamics.} The mean square displacement $\langle d^2\rangle(t)$ appears to have the following scaling behaviour
\begin{eqnarray}
\label{x2t}
\langle d^2\rangle(t)  \approx  A(\phi,T) t^{\eta_2} \qquad &&\text {if $t\ll t_\xi$} \nonumber \\
\langle d^2\rangle(t)   \approx   D(\phi,T) t \qquad &&\text {if $t\gg t_\xi$}  .
\end{eqnarray}
where the notation for the cross-over time $t_\xi$ will become clear below. It appears that the exponent $\eta_2$ is strictly less than unity. In fact below we show that there exists a fractal dimension (of the diffusive process) $d_w >2$ such that $\eta_2 = 2/d_w$.

The long-time asymptotic motion is diffusive with a diffusion coefficient $D(\phi,T)$. Diffusion  is strongly suppressed as the volume fraction approaches the jamming transition $\phi \rightarrow \phi_J$. The crossover time $t_\xi(\phi,T)$ can be expected to diverge as the jamming transition is approached.  This behaviour can be seen in the upper panel of Fig.~\ref{msdvt} and we will estimate $t_\xi(\phi,T)$ as well as all exponents from scaling arguments.
\begin{figure}
\includegraphics[scale=0.22]{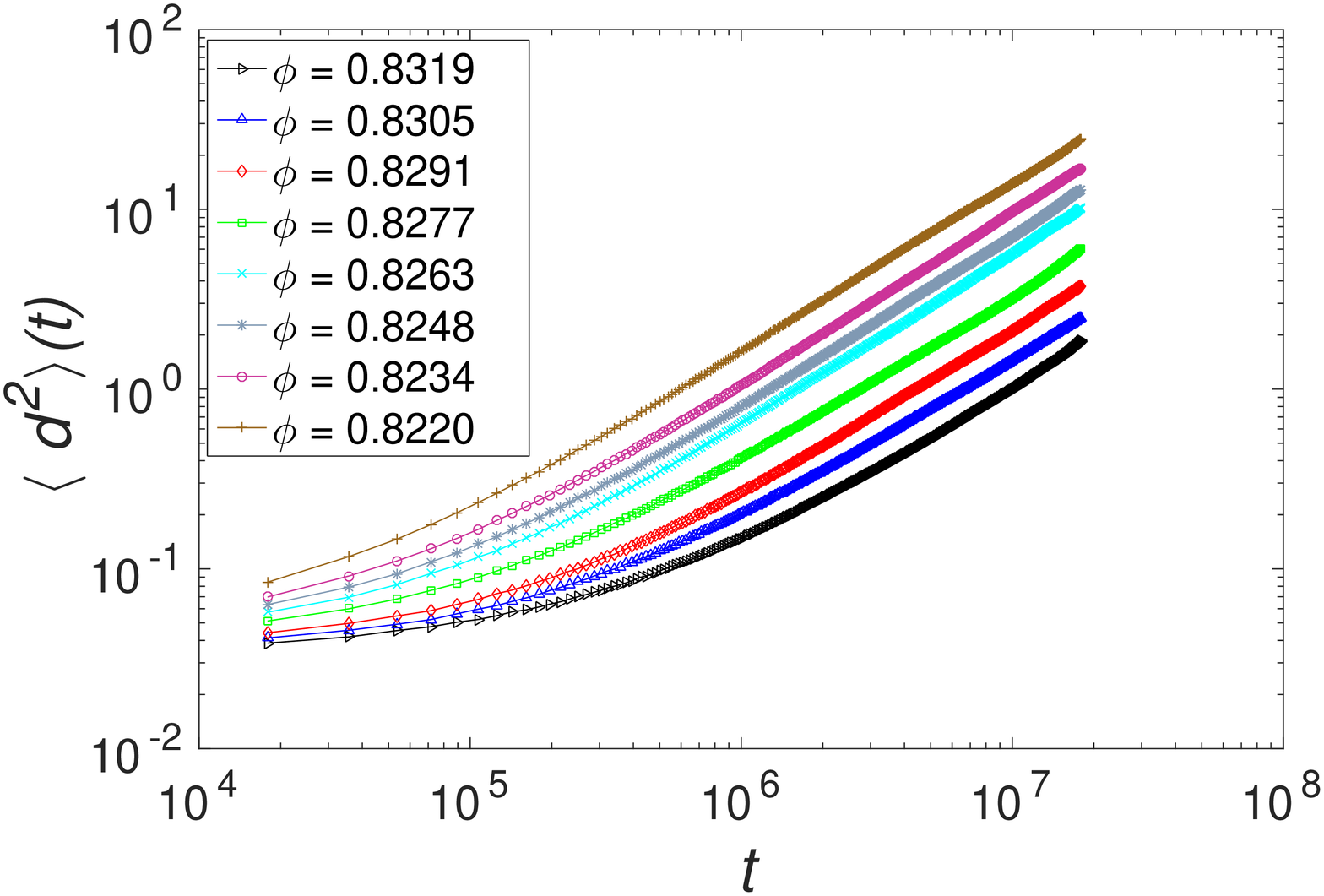}
\vskip 0.5 cm
\includegraphics[scale=0.22]{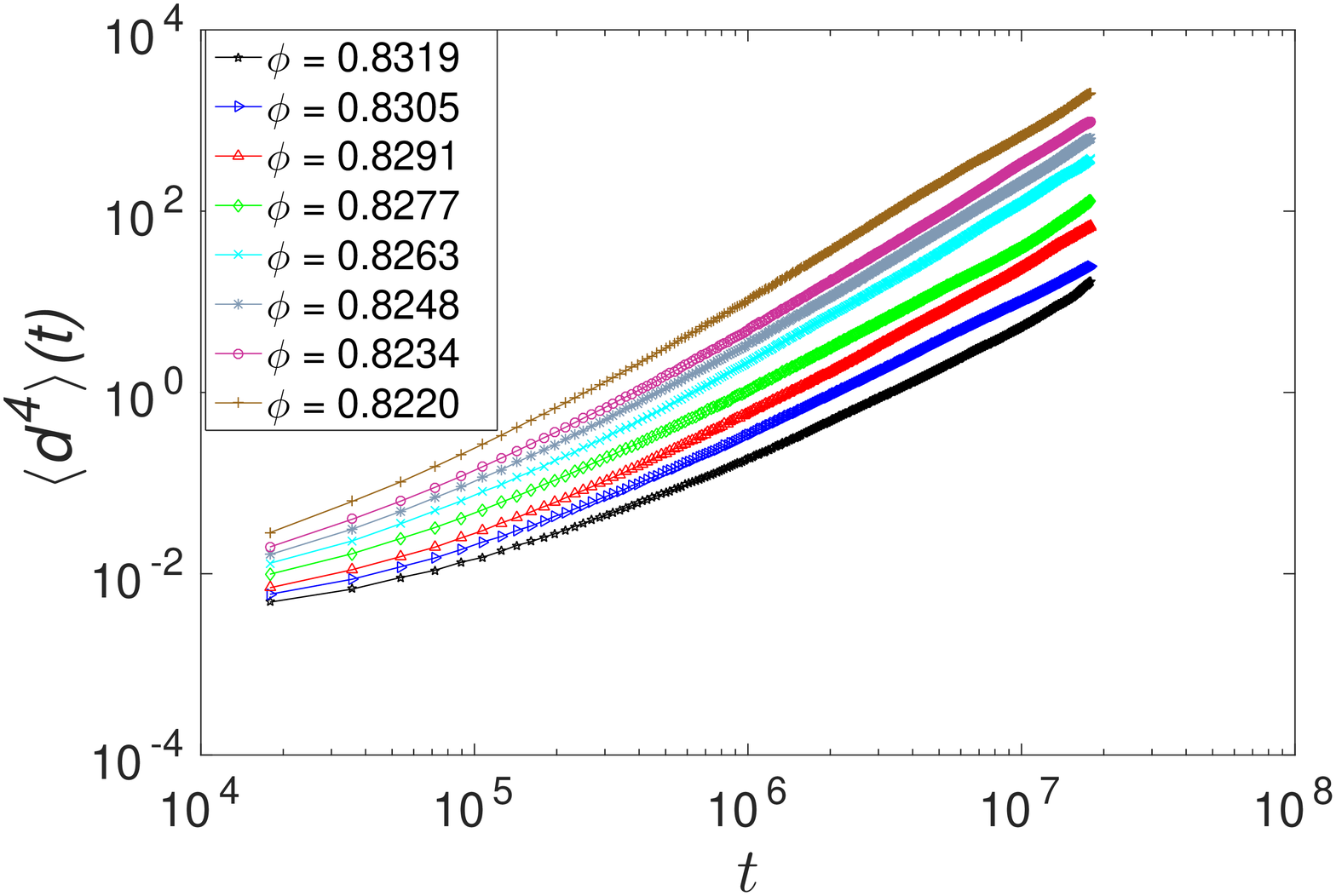}
\caption{Upper panel: Mean square displacement versus time for different $\phi_J-\phi$ and $T=1.167$ .
Lower panel: The fourth moment of the displacement for different $\phi_J-\phi$ and $T=1.167$.}
\label{msdvt}
\end{figure}
\item {\bf The grain dynamics does not have a Gaussian PDF }. The motion of the grains, though apparently diffusive in nature, does not give rise to a Gaussian probability distribution. In other words,
     \begin{eqnarray}
     P(d_i,t) &\ne& \frac{1}{\sqrt{4\pi Dt}}\exp{[-\frac{d_i^2}{4Dt}]} \ .
     \label{standard}
     \end{eqnarray}
      The distribution is more similar to a stretched exponential (cf. upper panel of Fig.~\ref{figpdf}), and thus  multiscaling emerges naturally as a feature of granular diffusion in dense granular media close to jamming. This is shown in the lower panel of Fig.~\ref{msdvt} which displays the time dependence of the fourth order moment of the displacement:
\begin{equation}
\label{multi}
\langle d^4\rangle(t)   \approx  A_n t^x \quad \text {if $t\gg t_\xi$} \ .
\end{equation}
We find that $x\ne 2$. In fact in the diffusive regime at all packing
fractions, $x$ ranges in value from 1.5 at high values of $\phi$ to 1.8 at lower values.
The probability distribution functions are isotropic, however, as expected for walks which have no well defined drift in any particular direction (see lower panel Fig.~\ref{figpdf}).

\begin{figure}
\includegraphics[scale=0.25]{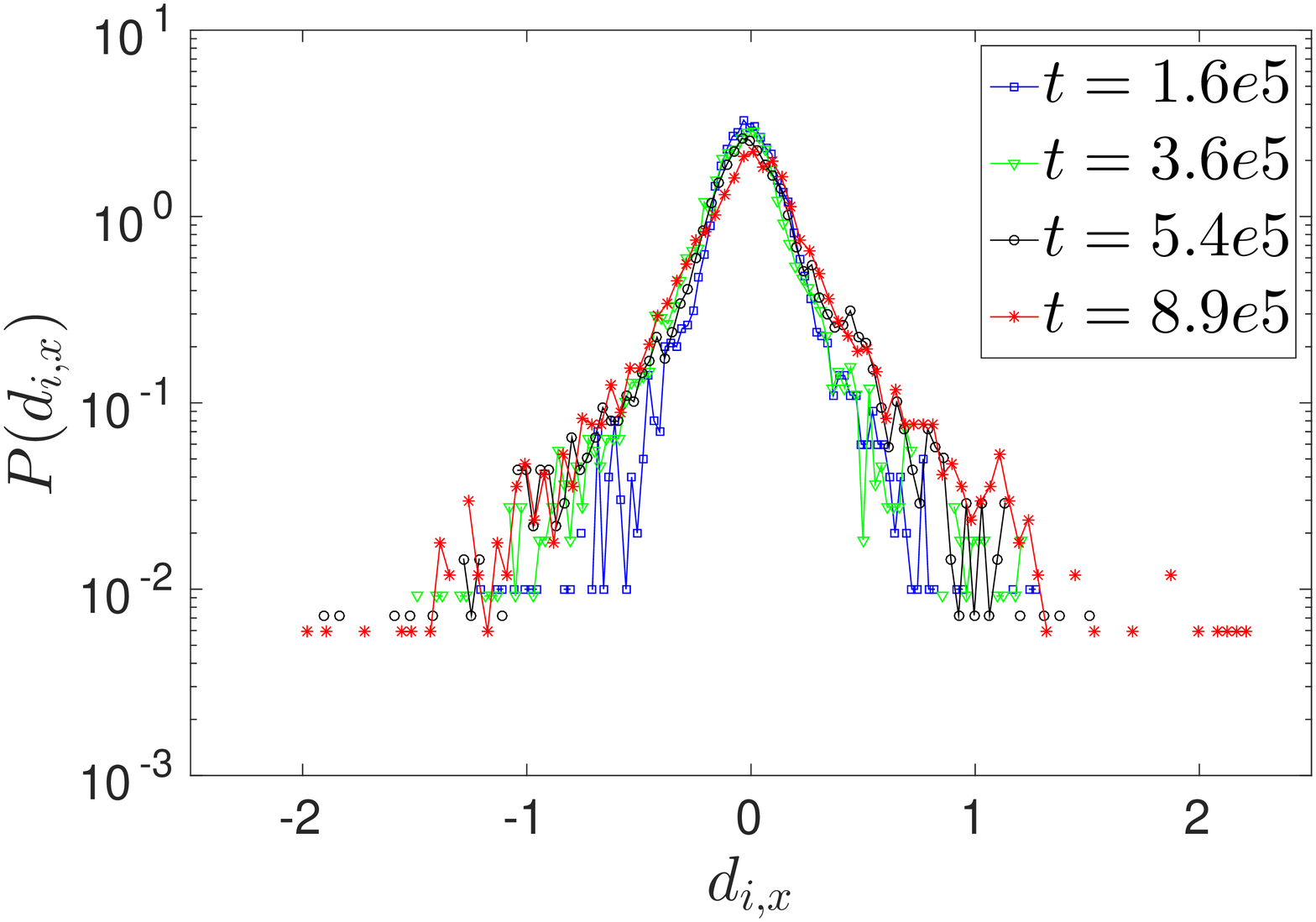}
\includegraphics[scale=0.20]{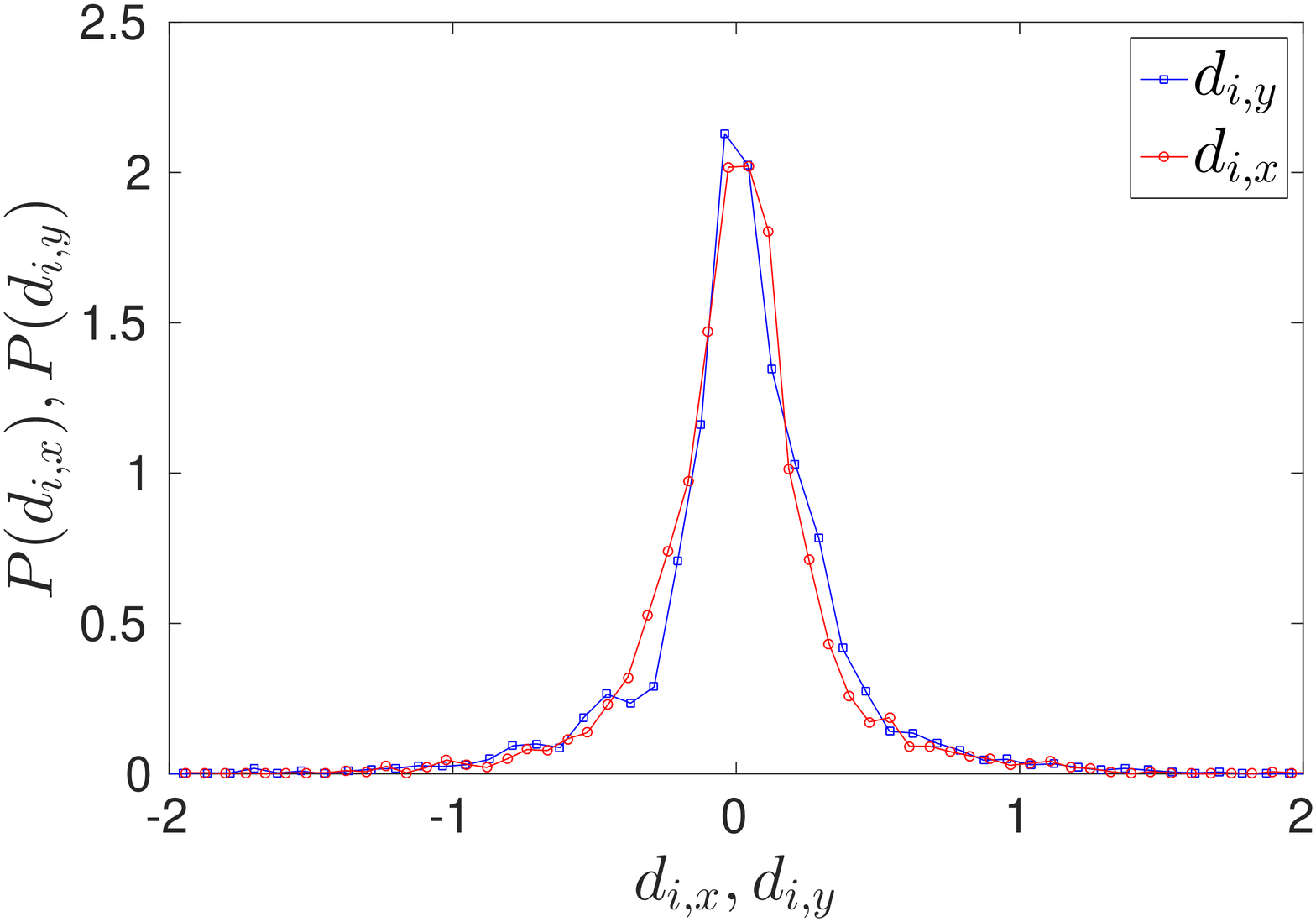}
\caption{Upper Panel: Logarithm of the probability distribution function $P(d_{i,x},t)$ plotted against distance $d_{i,x}$ for several times $t$. Note the non gaussian nature of the pdfs.
Lower Panel: The probability distribution function $P(d_{i,x},t)$ and $P(d_{i,y},t)$ plotted against distance $d_{i,x}$ and $d_{i,y}$ superimposed. The isotropy of the motion is clear.}
\label{figpdf}
\end{figure}
\end{itemize}

\section{Scaling of Diffusion in Granular Media}
\label{scaling}
In this section we develop a scaling approach to explain the data accumulated in the numerical simulations.
The final aim is to achieve a data collapse over the range of employed parameters. Such data collapse
offers a predictive theory in the sense that measuring the diffusive dynamics for one set of parameters allows
us to predict the same for any other set of parameters.

In identifying the relevant scales employed in the scaling theory we will refer below to published
results on the jamming transition in athermal (and frictionless) conditions. While it is obvious that
the numerical value of $\phi_J$ depends on friction and also its very existence is obscured by the
agitation, we will find that this approach is adequate for getting correct orders of magnitude and to
achieve data collapse. To begin, we recall that diffusion is always possible below $\phi_J$. In agitated granular media there is also the possibility of diffusion in a regime $\phi_J < \phi < \phi^*(T)$.  Even for
$\phi > \phi_J$ particles are able to diffuse if perturbed strongly enough. Let us first consider diffusion for $\phi \ll \phi_J$.

\subsection{Diffusion in a Granular Gas}
\label{lowdens}

In this subsection we consider diffusion in the granular gas phase $\phi \ll \phi_J$.  Denoting by $\bar\sigma$  the average diameter of a single grain, this regime of granular gas is
characterized by an average distance $\ell\gg \bar \sigma$ that separates the grains. Moreover, in this regime we estimate
$\ell^2 N\approx L^2$ or $\ell\sim L/\sqrt{N}$. Finally we note that $\phi\approx N\bar\sigma^2/L^2$ and therefore $\ell\approx \bar\sigma/\sqrt{\phi}$. Using this we then estimate the diffusivity $D$ of a single grain as
\begin{equation}
\label{diff}
D\sim \ell^2/t \sim v \ell \approx \sqrt{\langle v^2\rangle} (\bar\sigma/\sqrt{\phi}) \approx \sqrt{T/(m \phi)}\bar\sigma .
\end{equation}

Note that this granular temperature $T$ also allows us to define a typical time $\tau(T)$ as
\begin{equation}
\label{tau}
\tau(T) = \bar\sigma/v = \bar\sigma/\sqrt{T/m}\sim T^{-1/2} .
\end{equation}
This time is interpreted as a typical ``attempt time".

\subsection{Scaling Approach to Diffusion in Compact Granular Media Below Jamming}
\subsubsection{The energy barrier}
The temperature dependence for the diffusion constant $D(\phi, T) \sim \sqrt{T}$, which is  valid in the granular gas phase, must breakdown as $\phi \rightarrow \phi_J$ from below. When the area fraction increases, in order to make an actual hop a grain must overcome some energy barrier $\Delta(\phi)$. Thus we expect
in general
\begin{equation}
\label{tauT}
\tau_{\rm hop} (\phi,T) = \tau(T) \exp({\Delta(\phi)/T}) \approx \frac{\bar\sigma}{\sqrt{T/m}} \exp({\Delta(\phi)/T})\  .
\end{equation}
Note that this expectation has to be justified a-posteriori (as we do below) since our system is not
a standard thermal ensemble. To use this expression in our scaling arguments below we need to estimate how $\Delta$ depends on $\phi$.
When a disk hops from one position to another it interacts elastically with a number $\C N$ of other disks in its neighborhood.
The value of $\C N$ will be estimated below using the results of the simulations. With the addition of the
new disk the local area fraction changes from $\phi$ to $\phi'\approx (\C N+1) \phi/\C N$.  Before the hop the grains occupy an area $\C A \approx \C N \sigma^2/\phi$ and if $\phi <\phi_J$ the local pressure can be estimated from the ideal gas relation, i.e. $P\approx \phi T/\sigma^2$.  After the grain hops it creates a neighborhood of about $\C N + 1$ grains in the same area and the local area fraction is $\phi' > \phi$. Assuming that also $\phi'>\phi_J$ the disks are now suffering an average compression $\tilde \delta$ (cf. Eq.~(\ref{Fn}). Then the local pressure in this region becomes  $P'\approx \phi' T/\sigma^2 + K_n {\tilde \delta}^{3/2}/\sigma$ for Hertzian discs. This estimate is valid
as long as our area fraction $\phi$ was close enough to $\phi_J$. Then the local change in pressure due to a hop can be estimated as
\begin{equation}
\Delta P = P'-P = \frac{T(\phi'-\phi)}{\sigma^2} + \frac{K_n}{\sigma} {\tilde \delta}^{3/2} \ .
\label{pprime}
\end{equation}

We now need to estimate $\tilde\delta$.  To this aim we will use previous results pertaining to the jamming transition in frictionless soft spheres with Hertzian normal forces, asserting that estimates of pressure should depend mostly on the normal forces. Simulations of ensembles of Hertzian discs~\cite{03ODLN} suggest that just above the jamming transition at $T=0$ the pressure changes from $P=0$ to $P' \approx K_n \sigma^{1/2} (\phi'-\phi_J)^{3/2}$.
Comparison between this result and our expression (\ref{pprime}) for $\Delta P$ in terms of $\tilde \delta$ (for $T=0$) suggests that the overlap is linear with the change in area fraction above $\phi_J$
\begin{equation}
\label{overlap}
\tilde \delta  \approx \sigma (\phi'-\phi_J) \ .
\end{equation}
In consequence the typical energy barrier will scale as
\begin{eqnarray}
\label{deltaphi}
&&\Delta (\phi ) \sim \Delta P \sigma^2  \approx    T (\phi'-\phi) + K_n \sigma {\tilde \delta}^{3/2} \nonumber \\
&&\approx T \phi/\C N + K_n \sigma^{5/2} [(\C N+1) \phi/\C N - \phi_J]^{3/2}
\end{eqnarray}
Eq.~(\ref{deltaphi}) implies that at $\phi=\phi_J$ the energy barrier can be estimated as
\begin{equation}
\label{deltax}
\Delta_J = \Delta(\phi_J)  \approx  T\phi_J/\C N + K_n \sigma^{5/2}(\phi_J/\C N)^{3/2}  .
\end{equation}
and thus close to jamming and at low temperatures ($T \ll \Delta_J$ ) the energy has a linear dependence on the area fraction
\begin{equation}
\label{delta2}
\Delta (\phi) /\Delta_J  \approx 1+ C (\phi-\phi_J)/\phi_J  ,
\end{equation}
where $C \approx 3(\C N +1)/2$. The value of $C$ will be determined below from simulations, vindicating
the functional form Eq.~(\ref{delta2}). We should stress here that we assume explicitly that $\Delta (\phi)$
is independent of $T$. Eq.~(\ref{delta2}) indicates that this assumption requires $\phi_J$ to be independent
of $T$. This cannot be exact. Nevertheless we do find below an excellent fit with $T$-independent $\phi_J\approx 0.838$. This approximation limits how close we can approach jamming, and see the discussion section for more
details.

We will see below that the data support very well the assumption of $T$-independent $\Delta (\phi)$. One
can speculate that the physical reason for this is that the thermal pressure is negligible as compared to the mechanical pressure, once a grain attempts to increase locally the packing fraction above $\phi_J$ \cite{Olivier}. Our estimate below for $\Delta_J$ is of the order of 10, and our temperatures are always obeying the constraint
$T \ll \Delta_J$.

\subsubsection{The typical lengthscale}
To proceed we assume that there exists a length scale
\begin{equation}
\label{xi}
\xi (\phi) \approx \sigma (\phi_J - \phi)^{-\nu},
\end{equation}
that diverges as $\phi \rightarrow \phi_J$ from below with a power law form. This is in agreement with the measurement in Refs.~\cite{05DHRR,07KAGD} of typical scales. These can be either the influence length (the total number of moving disks) as a single disk is pulled through the medium (and see also \cite{12FGZ}, or a correlation length as measured in Ref.~\cite{07KAGD}. These scales diverge as indicated in Eq.~(\ref{xi}), with the measured exponent in two dimensions lying between 0.4 (\cite{07KAGD}) and 0.7 (\cite{05DHRR}).
We assume that this lengthscale is also the typical distance between nucleating sites, $l_{\rm nuc}(\phi,T )\sim \xi$, when the temperature is low enough. We will use $\xi$ as an order of magnitude
estimate of $l_{\rm nuc}(\phi,T )$ remembering that for high $T$, $l_{\rm nuc}(\phi,T )$ will become significantly smaller than $\xi$.
Then we can estimate the crossover time $t_{\xi}$ as the time it takes particles to diffuse between neighbouring nucleating sites.

Using Eq.~(\ref{x2t}) at crossover time we write
\begin{eqnarray}
\label{xi2t}
\left(\frac{\xi}{\sigma}\right)^2  & \sim &  \left(\frac{t_{\xi}}{\tau_{\rm hop}}\right)^{2/d_w}  \nonumber \\
\xi^2  & \sim &  D t_{\xi}   .
\end{eqnarray}
Here $d_w$ is the fractal dimension of the trajectory, in other words $t/\tau\sim \left(r/\sigma\right)^{d_w}$.
Thus we see first that
\begin{equation}
\label{txi}
\frac{t_{\xi}}{\tau_{\rm hop}} \sim\left(\frac{\xi}{\sigma}\right)^{d_w}
\end{equation}
and secondly
\begin{eqnarray}
\label{dxi}
D &\sim& \frac{\xi^2}{t_{\xi}} \sim \frac{\sigma^2}{\tau_{\rm hop}}\left(\frac{\xi}{\sigma}\right)^{2-d_w}
\nonumber\\&\sim&  \frac{\sigma^2}{\tau(T)} (\phi_J - \phi)^{\nu (d_w-2)}.
\end{eqnarray}
Denoting $d_w = 2 + \theta$ we can also rewrite Eq.~(\ref{dxi}) in the form
\begin{eqnarray}
\label{dxib}
D(\phi,T)  &\sim&  \frac{\sigma^2}{\tau(T)} (\phi_J - \phi)^{\nu \theta} \\ &\sim&  (\sigma \sqrt{T/m}) \exp{\left(\frac{-\Delta(\phi)}{T}\right)} (\phi_J - \phi)^{\nu \theta} \ . \nonumber
\end{eqnarray}
This is an important intermediate result and we check it against simulations.

To test the result we need first to measure the function $\Delta(\phi)$. To this aim we consider our data for $D(\phi, T)$
for fixed temperature $T$ and for all the available values of $\phi$. An example of the determination of
$\Delta(\phi)$ for a given value of $\phi$ is shown in the upper panel of Fig.~\ref{Deltaphi}. Once we have extracted $\Delta(\phi)$ for different values of $\phi$ we can test
the predicted scaling Eq.~(\ref{delta2}) with
$(\phi_J - \phi)$. The excellent agreement with our assumption can be seen in lower panel of Fig.~\ref{Deltaphi}.
Note that the numerical value of $\phi_J\approx 0.838$  is chosen here to get a best fit and used throughout the analysis.

\subsubsection{The fractal dimension $d_w$.}
\label{dw}

To understand the exponent $d_w\approx 4.0$ we use again scaling arguments. Consider the short time dynamics where diffusing disks cover a ball of radius $R(t)$ such that $\sigma\ll R(t)\ll \xi$. This region contains $(R(t)/\sigma)^2$ disks which most of them are not moving at all, but there exist a small and finite number of disks, say, $M$, that are diffusing appreciably, covering an area of the order of $\sigma^2$ in a time that is
of the order of $\tau_{\rm hop}$. In a longer time $t$, they are covering an area  $[\sigma^2/\tau_{\rm hop}(T)] t$. Thus the average mean square displacement is of the order of
\begin{equation}
 R^2(t) \sim  \frac{\sigma^2}{\tau_{\rm hop}} t \frac{M}{(R(t)/\sigma)^2} \ .
\end{equation}
Since $M$ is finite, we can estimate
\begin{equation}
\label{dxib2}
\frac{R(t)}{\sigma} \approx \left(\frac{t}{\tau_{\rm hop}}\right)^{1/4}.
\end{equation}
Thus $d_w \approx 4.0$, and $\theta \approx 2.0$. Note that this argument were valid all dimensions $d$ then  $d_w = 2+d$.

\section{Scaling functions}
\label{compare}

In this section we reap the benefit of the scaling relations derived above and continue to
derive a scaling form for the mean square displacement and the probability distribution function (pdf) of displacements. We should be able to express all quantities of interest in terms of $\xi$ and $t_\xi$. For example the mean square displacement is written as
\begin{equation}
\label{d2tscal}
\frac{\langle d^2\rangle(t)}{\xi^2}= G\left(\frac{t}{t_{\xi}}\right) \ .
\end{equation}
where the scaling function $G(y)$ has the asymptotic forms which are gleaned from Eq.~(\ref{x2t}):
\begin{eqnarray}
\label{g}
G(y) & \sim& y^{2/d_w} \quad \text{ for }~ y\ll 1\ ,  \nonumber \\
G(y) & \sim & y \quad \text {for }~y\gg 1 \ .
\end{eqnarray}
The meaning of this result is that we can predict $\langle d^2\rangle$
at any value of the parameters from the measurement of this quantity at
any given value of these parameters.

Next we consider the scaling function associated with the pdf $P(d_i,t)$. In an isotropic
system this pdf is independent of $i$ and we expect it to be a scaling function,
\begin{equation}
\label{pdf1}
P(d_{i,x},t) = \frac{1}{\sqrt{\langle d_{i,x}^2\rangle(t)}} f\left( \frac{d_{i,x}}{\sqrt{\langle
d_{i,x}^2\rangle(t)}} \right) \ .
\end{equation}
The reader should note that this scaling form differs in two important aspects from the
standard solution of the diffusion equation Eq.~(\ref{standard}). First, the function $f$ is
not a Gaussian, and second, $\langle d_i^2\rangle(t)$ is not diffusive except maybe at long times.
We will test the scaling form in the next section.

\section{Extracting scales and exponents from the numerics}
\label{extract}
\begin{figure}
\includegraphics[scale=0.20]{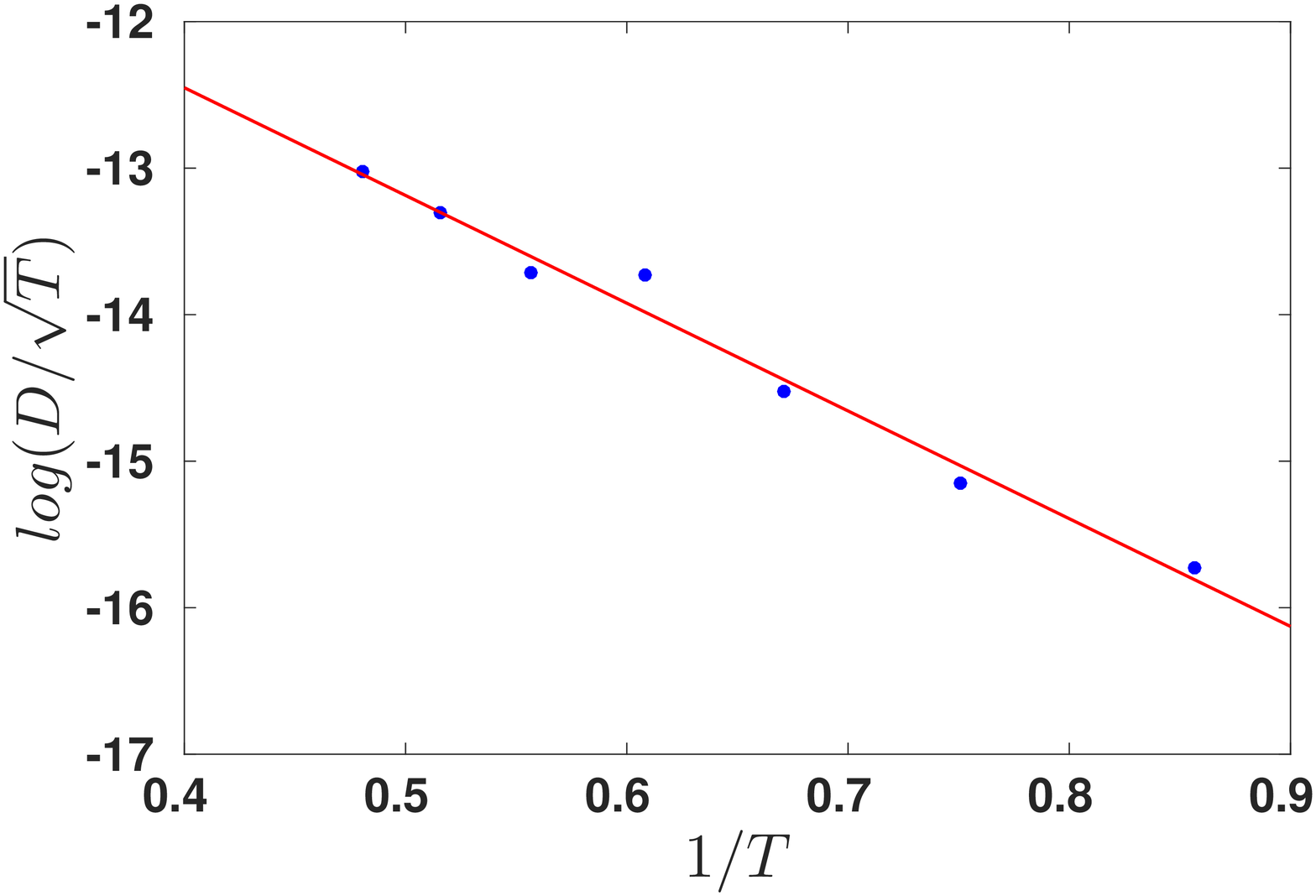}
\includegraphics[scale=0.20]{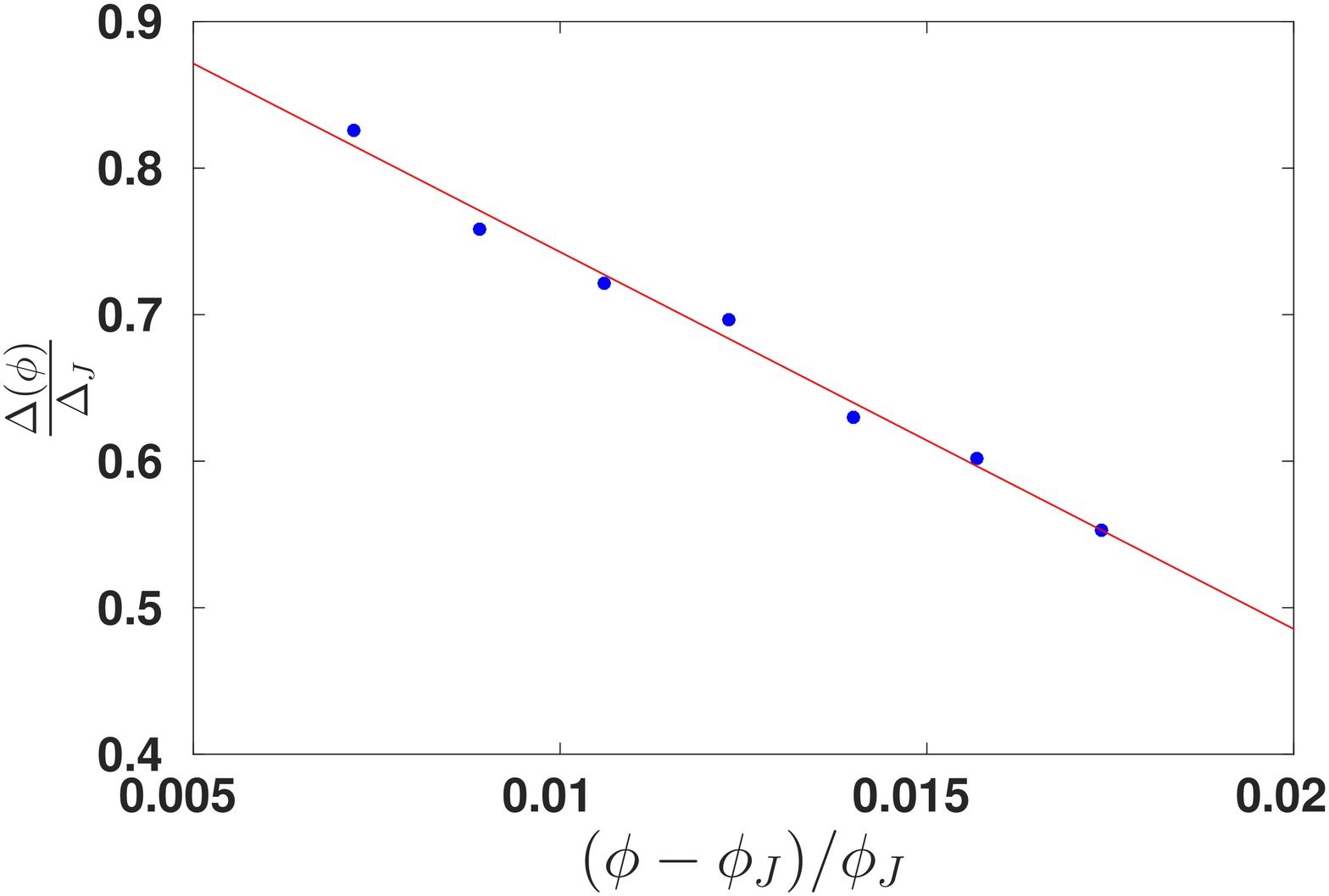}
\caption{Upper panel: an example of the extraction of $\Delta(\phi)$ for $\phi=0.832$, see text for details.
Lower panel: The data obtained for $\Delta(\phi)$ re-plotted as suggested by Eq.~(\ref{delta2}). The
agreement with the linear prediction should be noted.}
\label{Deltaphi}
\end{figure}

We begin by estimating the value of  $\C N$ using Eq.~(\ref{delta2}). We read the slope of the plot in Fig.~\ref {Deltaphi} lower panel, finding a value of 25.72. From Eq.~(\ref{delta2}) we then estimate $\C N \approx 16$. This is a very reasonable number which indicates that the hop of one disk influences appreciably all the nearest and next-nearest neighbors.
\begin{figure}
\includegraphics[scale=0.25]{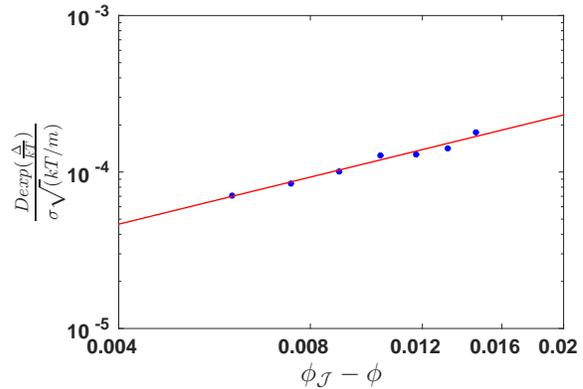}
\caption{An example of the extraction of the exponent $\nu$. This is done using Eq.~(\ref{dxib}) as explained
in the text. The present example pertains to $T=1.94$ and it results with the estimate $\nu\approx 0.5$.}
\label{fixnu}
\end{figure}

Once we have a good fit for $\Delta(\phi)$ we can return to Eq.~(\ref{dxib}) and re-plot
$D/[ (\sigma \sqrt{T/m}) \exp{\left(\frac{-\Delta(\phi)}{T}\right)} ]$ in a log-log plot as a function
of $\phi-\phi_J$ to extract the value of the exponent $\nu$. An example of this procedure for $T=1.94$
is shown in Fig.~\ref{fixnu}. Of course, every value of $T$ will give us a slightly different value
of $\nu$, but all the obtained values are in the narrow range $\nu=0.48\pm 0.024$. Of course one cannot
exclude the possibility that $\nu=1/2$. A discussion of this result and a comparison with the measurement of the typical scale in Ref.~\cite{05DHRR} is offered in the next section.

\begin{figure}2
\includegraphics[scale=0.20]{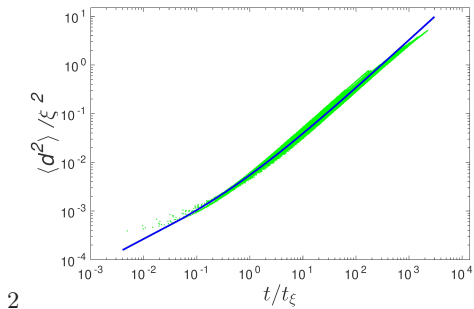}
\caption{A test of the scaling theory. The second moment of displacements $\langle d^2\rangle(\phi,T)$ for all our measured temperatures and all the values
of $\phi$ re-scaled according to the theoretical prediction Eq.~(\ref{d2tscal}). The data collapse
is a clear indication of the success of the scaling theory.}
\label{success}
\end{figure}

Now we are ready for the major test of our scaling approach. We should go back to the data of the type shown
in the upper panel of Fig.~\ref{msdvt}, but for all the available temperatures and area fractions, and re-plot the data according to Eq.~(\ref{d2tscal}). At this point
we have no free scale and no free exponent, so a good data collapse will serve as a strong support
to the scaling approach advocated above. Indeed, the re-scaled data as shown in Fig.~\ref{success} is
extremely satisfactory, indicating that we identified the right scales and reasonable exponents.

\begin{figure}
\includegraphics[scale=0.24]{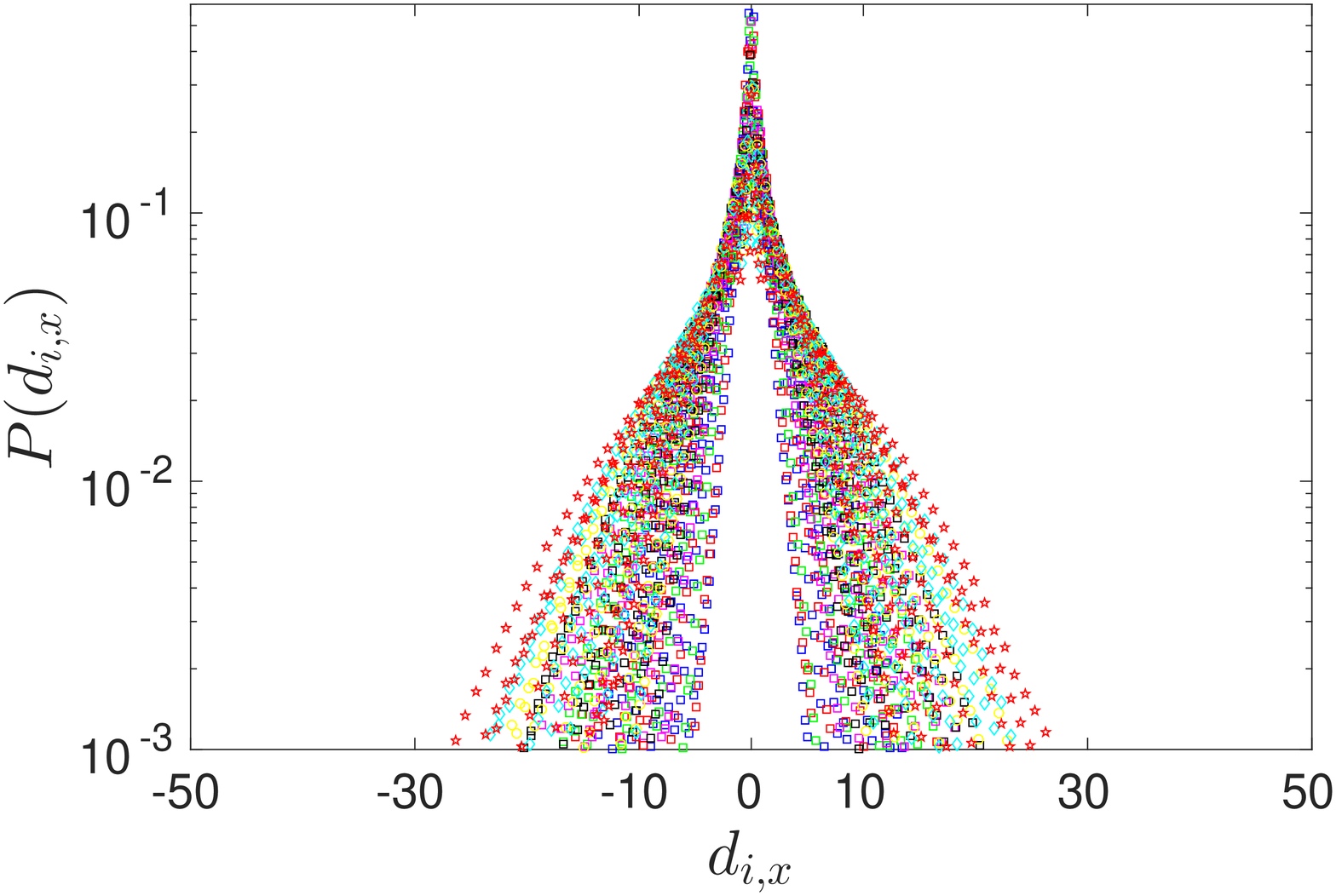}
\vskip 0.1 cm
\includegraphics[scale=0.24]{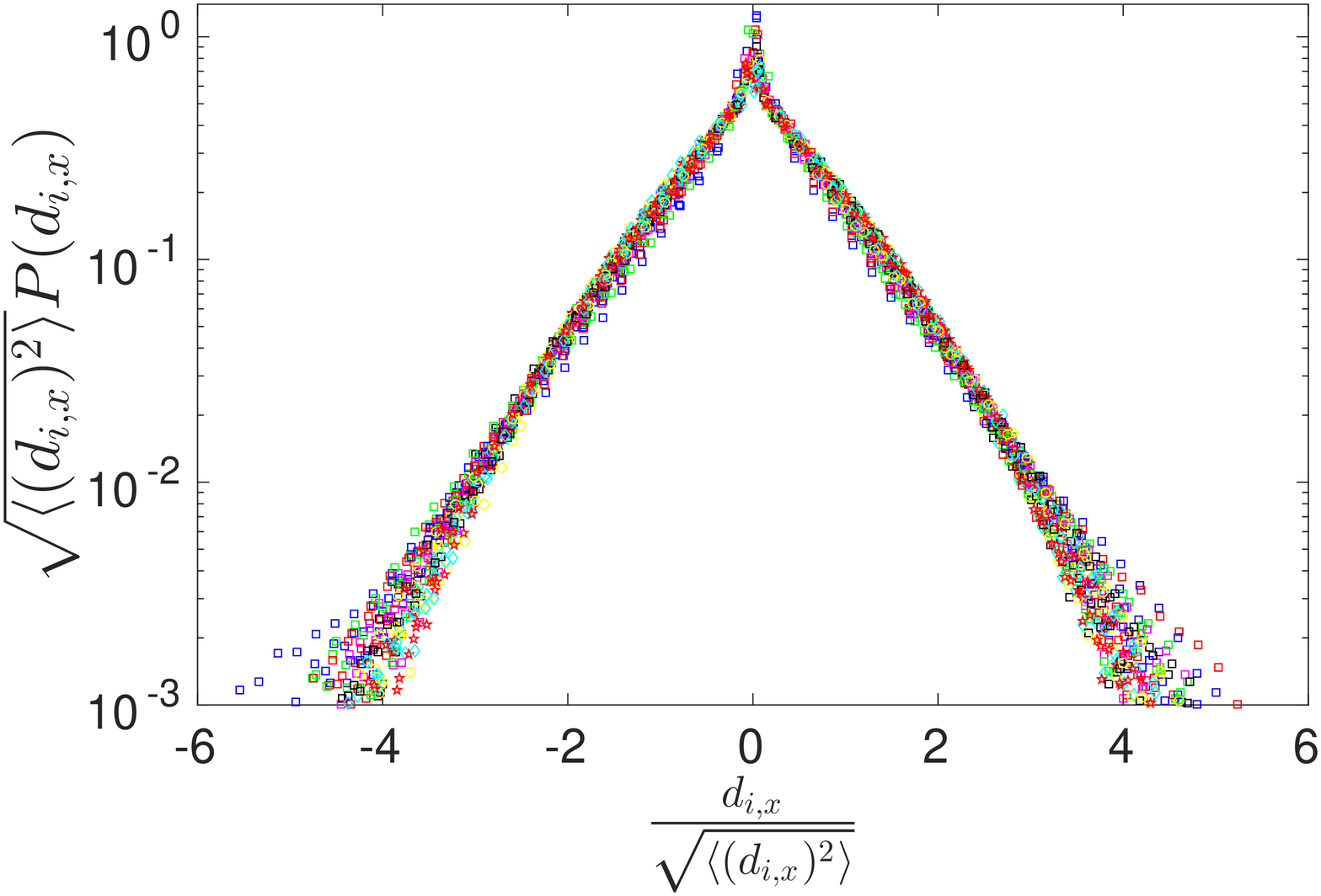}
\caption{ Upper panel: A typical example of the pdf $P(d_{i,x})$ for a given time in the diffusive
regime for different values of $\phi\in [0.822,0.832]$ and $T\in [1.16,2.08]]$. The pdf's broaden
when $T$ increases and $\phi$ decreases. Lower panel: the data collapse upon re-scaling according to
Eq.~(\ref{pdf1}). }
\label{testpdf}
\end{figure}

Finally, we test the prediction of scaling theory for the pdf functions Eq.~(\ref{pdf1}). Directly measured pdf functions (without re-scaling) are shown in the upper panel of Fig.~\ref{testpdf}. For probabilities smaller than $10^{-3}$ the data become noisy due to paucity of sufficiently active walkers. The reader should note that these pdf's are very far from Gaussian form, underlying the fact that the diffusive behavior of the second moment of
displacement should not be confused with ``simple" diffusion.

The data collapse obtained
by re-plotting according to Eq.~(\ref{pdf1}) is shown in the lower panel. The reader can conclude that the scaling theory appears vindicated.

\section{Discussion and conclusions}
\label{discussion}

A very large body of work has been devoted over the last couple of decades to the jamming transition, with
stress at the athermal scaling properties near $\phi_J(T=0)$ and especially above the transition at $\phi > \phi_J$. For the agitated systems studied in this paper, the notion of jamming at $\phi_J$ becomes fuzzy, since
diffusion can continue if the agitation is sufficiently vigorous. Note for example that
Eq.~(\ref{dxib}) taken literally predicts that $D(\phi_J,T)=0$ for any $T$. This is of course incorrect
since our $\Delta(\phi_J)$ is finite. This should serve to underline the fact that in this paper we
avoid the immediate vicinity of $\phi_J$ and that there must exist some sort of crossover to the regime studied in Ref.~\cite{08LDBB}. We find that in the regime that we explore we can use in the scaling theory a value of $\phi_J$ which is independent of $T$. It is likely that this stems from the fact that the thermal pressure is negligible as compared to the mechanical pressure, once a grain attempts to increase locally the packing fraction above $\phi_J$. Another theoretical issue that needs attention is the role of the ``temperature" $T$ in the scaling approach. We explained above that our system is open, with constant agitation and dissipation. Thus fluctuation dissipation theorems are not expected to hold. Nevertheless in Eq.~(\ref{tauT}) we have assumed that
an Arrhenius form with $T$ interpreted as $m\langle v^2\rangle $ represents barrier crossings with a barrier
denoted above as $\Delta(\phi)$. Such assumptions can and must be tested a-posteriori as we did in Sect.~\ref{extract}. The conclusion is that for analyzing the transport properties of agitated assemblies of frictional Hertzian disks,
scaling ideas appear relevant and useful below the jamming transition and even
at some sizeable distance from athermal conditions. The central quantities that enable the scaling theory
were the typical scale $\xi$, and the typical time-scale $t_\xi$. In discussing $\xi$ we identified this
typical scale with $l_{\rm nuc}$ that was introduced in Sect.~\ref{simresults}. In our numerical fits we found a scaling exponents $\nu\approx 0.5$ with which we get very good data collapse. Comparing with the result of
Refs.~\cite{05DHRR,07KAGD} in which the exponent ranged between 0.4 and 0.7, we cannot say at present whether all these length scales are the same or not. We stress however that $l_{\rm nuc}$ is a function of both $\phi$ and $T$, and in fact it is a monotonically decreasing function of $T$. It is possible that as $T\to 0$ (a regime not covered in this paper) the similarity of all these scales becomes more evident.

In contrast, the exponent $d_w$, which sets the temporal scale could be estimated theoretically (cf. section \ref{dw}). We used the theoretical result $d_w=4$ throughout the numerical fits with excellent data collapse.

As said, for temperatures high enough diffusion continues also at $\phi>\phi_J(T=0)$ .
We did not study such conditions in the present paper and it may appear useful to concentrate on this regime in future studies, to achieve complete understanding of the transport properties over the whole range of area fraction and temperature.

\acknowledgments
This work has been supported in part by the US-Israel BSF, the Israel Science Foundation under the joint program with Singapore and the IMOS collaborative program with Italy. We are grateful to Olivier Dauchot for proposing the problem to us and for his critical reading of the first draft.

\bibliography{ALL}

\end{document}